\documentclass[acmtog,authorversion]{acmart}

\usepackage{booktabs} 

\definecolor{cb-black}      {RGB}{  0,   0,   0}
\definecolor{cb-blue-green} {RGB}{  0,  073,  073}
\definecolor{cb-green-sea}  {RGB}{  0, 146, 146}
\definecolor{cb-rose}       {RGB}{255, 109, 182}
\definecolor{cb-salmon-pink}{RGB}{255, 182, 119}
\definecolor{cb-purple}     {RGB}{ 73,   0, 146}
\definecolor{cb-blue}       {RGB}{ 0, 109, 219}
\definecolor{cb-lilac}      {RGB}{182, 109, 255}
\definecolor{cb-blue-sky}   {RGB}{109, 182, 255}
\definecolor{cb-blue-light} {RGB}{182, 219, 255}
\definecolor{cb-burgundy}   {RGB}{146,   0,   0}
\definecolor{cb-brown}      {RGB}{146,  73,   0}
\definecolor{cb-clay}       {RGB}{219, 209,   0}
\definecolor{cb-green-lime} {RGB}{ 36, 255,  36}
\definecolor{cb-yellow}     {RGB}{255, 255, 109}

\usepackage[ruled]{algorithm2e} 

\SetAlFnt{\small}
\SetAlCapFnt{\small}
\SetAlCapNameFnt{\small}
\SetAlCapHSkip{0pt}

\setcopyright{none}

\definecolor{humanoidblue}{RGB}{92,148,252}

\begin{document}
\fancyfoot{}
\pagestyle{plain} 

\title{HIL: Hybrid Imitation Learning for Dynamic Athletic Control}

\author{Jiashun Wang}
\affiliation{%
  \institution{Carnegie Mellon University}
  \country{USA}}
\email{jiashunw@andrew.cmu.edu}

\author{Yifeng Jiang}
\affiliation{%
  \institution{NVIDIA}
  \country{USA}}
\email{yifengj@stanford.edu}

\author{Haotian Zhang}
\affiliation{%
  \institution{NVIDIA}
  \country{USA}}
\email{haotianz@nvidia.com}

\author{Chen Tessler}
\affiliation{%
  \institution{NVIDIA}
  \country{Israel}}
\email{ctessler@nvidia.com}

\author{Davis Rempe}
\affiliation{%
  \institution{NVIDIA}
  \country{USA}}
\email{drempe@nvidia.com}

\author{Jessica Hodgins}
\affiliation{%
  \institution{Carnegie Mellon University}
  \country{USA}}
\email{jkh@cs.cmu.edu}

\author{Xue Bin Peng}
\affiliation{%
  \institution{Simon Fraser University}
  \country{Canada}}
\affiliation{%
  \institution{NVIDIA}
  \country{Canada}}
\email{xbpeng@sfu.ca}

\begin{abstract}
Data-driven methods leveraging deep reinforcement learning have become the dominant paradigm for developing controllers that enable physically simulated characters to produce natural human-like behaviors. However, these data-driven methods often struggle to adapt to novel environments and compose diverse skills to perform more complex interaction tasks with the environment. To address these challenges, we propose a hybrid imitation learning (HIL) framework that combines motion tracking, for precise skill replication, with adversarial imitation learning, to enhance adaptability and skill composition, enabling robust dynamic control for highly athletic behaviors. This hybrid learning framework is implemented through parallel multi-task environments and a unified observation space, utilizing a goal-conditioned representation to facilitate knowledge-sharing across the hybrid parallel environments. 
We demonstrate the effectiveness of HIL on a parkour-style obstacle traversal task and a heading control task. Our framework enables a unified controller that not only preserves the naturalness of reference motion data, but also generalizes effectively to challenging new environments. Evaluations across procedurally generated tasks and baselines show that our method improves motion quality, increases skill diversity, and achieves competitive task completion compared to previous learning-based approaches. Results are best visualized through \textcolor{humanoidblue}{\url{https://jiashunwang.github.io/HIL}.}
\end{abstract}


%
%


\begin{teaserfigure}
\includegraphics[width=\textwidth]{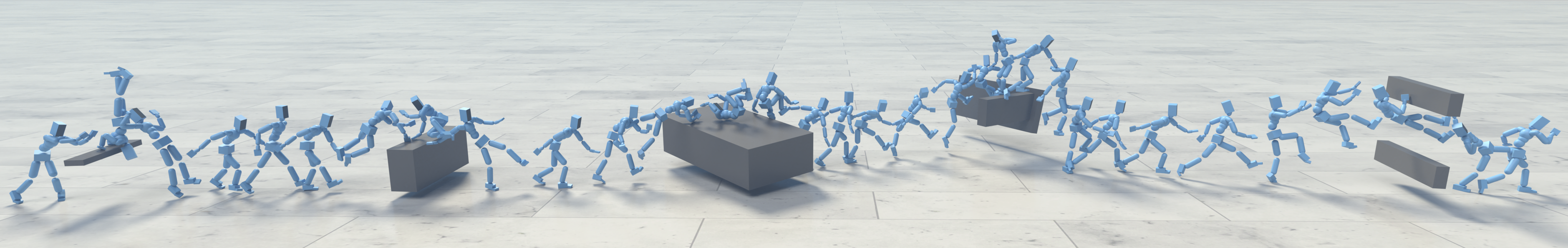}
\vspace{-0.2in}
\caption{We propose a Hybrid Imitation Learning (HIL) framework that is able to train a unified controller to master a diverse range of parkour skills and execute agile life-like interactions with various obstacles. In this example, a physically simulated character employs five distinct parkour skills to successfully clear the obstacle course.}
\label{fig:teaser}
\end{teaserfigure}

\maketitle

\section{Introduction}

Physically simulated characters that can replicate human behaviors have broad applications in animation, virtual reality, and robotics. These applications can benefit from enhanced physical realism, which increases user engagement in virtual settings, and improved task performance in robotics. Recent data-driven methods for physics-based character animation have made significant progress in generating agile behaviors, such as running, boxing, and tennis. By leveraging deep reinforcement learning (DRL)~\cite{DBLP:journals/tog/PengALP18, DBLP:journals/tog/LiuH18}, which simplifies the design of control structures, these methods facilitate the development of simulated characters that are both visually and functionally realistic.

While recent methods show promise, a key challenge remains: developing a unified control policy that can adapt to new scenarios with a diverse set of motor skills. Although motion tracking techniques can closely replicate a wide range of motor skills~\cite{DBLP:journals/tog/PengALP18}, they often lack the flexibility to adapt skills to novel environments, such as sequencing and composing different skills. Alternatively, general distribution matching techniques, such as adversarial imitation learning~\cite{DBLP:journals/tog/PengMALK21,DBLP:journals/corr/abs-2105-10066}, provide greater flexibility to modify and adapt skills to new scenarios. However, these methods are susceptible to mode collapse and may result in less natural and repetitive behaviors.
\textcolor{black}{These limitations are particularly evident in dynamic athletic tasks with complex scene interactions, such as parkour, where characters must sequence multiple dynamic stunts and adapt them to obstacles in the environment. In these scenarios, motion tracking approaches often struggle to generalize skills beyond the reference trajectories, while adversarial imitation learning methods may collapse to repetitive interaction strategies that ignore scene-specific affordances.}

To overcome these limitations, we introduce a simple but effective hybrid imitation learning method (HIL). HIL jointly trains a controller in two modes: (1) a motion tracking mode designed to closely replicate various parkour skills, and (2) an adversarial imitation learning mode designed to enhance smooth transitions between skills and adaptability across different scenarios. This hybrid framework is instantiated through parallel multi-task environments with a unified observation space shared by both modes. \textcolor{black}{This shared representation enables skills learned from reference data to transfer more effectively to goal-driven scenarios, while mitigating mode collapse. As a result, HIL is able to train a unified controller capable of composing diverse athletic skills and performing agile, life-like behaviors in novel and challenging environments.}

For tasks such as parkour, where aligned motion data with scene geometry is scarce, we source reference motion clips from online videos. For more common tasks, such as heading control, we leverage existing motion capture datasets. We systematically evaluate our model in a large number of diverse procedurally generated environments. We benchmark our method against multiple baseline techniques from prior work, and our results demonstrate improved motion quality, reduced mode collapse, and broader skill coverage.

\section{Related Work}

Data-driven methods have revolutionized character animation. With a comprehensive motion dataset, kinematic-based methods, such as motion graphs~\cite{DBLP:journals/tog/KovarGP02, DBLP:journals/tog/SafonovaH07} and motion matching~\cite{clavet2016motion}, can produce realistic animations. Later, deep learning models advanced motion synthesis~\cite{DBLP:journals/tog/HoldenKS17, DBLP:journals/tog/StarkeMK22}. However, these approaches struggle with dynamic interactions and generalization to unseen scenarios, motivating the exploration of physics-based techniques.

\textit{Physics-based models} simulate forces and dynamics to produce physically plausible behaviors. By leveraging reference motion clips, these methods can generate more realistic character movements~\cite{DBLP:conf/sca/ZordanH02,DBLP:journals/tog/ZordanMCF05,DBLP:journals/tog/SafonovaHP04,DBLP:journals/cgf/SilvaAP08, DBLP:journals/tog/LeeKL10, DBLP:journals/tog/LiuYPG12}. The strategic use of these reference motions has been a focal point of study. Motion tracking is a widely used technique for creating controllers that closely follow reference trajectories, typically by optimizing a tracking objective~\cite{DBLP:journals/tog/PengALP18, DBLP:journals/tog/SokKL07,DBLP:journals/tog/LiuPY16,DBLP:journals/tog/LiuH18,DBLP:journals/tog/FussellBH21, DBLP:journals/tog/WonGH20,DBLP:conf/iccv/0002CWKX23,DBLP:journals/tog/TesslerGNCP24, DBLP:journals/corr/abs-2502-20390}. However, simply mimicking reference motions is insufficient for adapting to new scenes. Kinematic planners can be used to synthesize or retrieve new references~\cite{DBLP:conf/nips/0007K20, DBLP:journals/tog/ParkRLLL19, DBLP:journals/tog/BergaminCHF19}. However, they often lack physical plausibility in new scenes that require a lot of interactions. Incorporating task objectives can help generate new behaviors~\cite{DBLP:journals/tog/PengALP18}, but frame-by-frame tracking rewards restrict necessary deviations from reference motion, thereby limiting the system's adaptability.

The integration of Adversarial Imitation Learning (AIL) marked a significant advancement~\cite{DBLP:conf/nips/HoE16,DBLP:journals/tog/PengMALK21,DBLP:journals/corr/abs-2105-10066}. AIL approaches learn the distribution of reference motions through a discriminator~\cite{DBLP:journals/tog/XuSZK23, DBLP:journals/tog/XuXAKNMKZ23, DBLP:conf/corl/LiVBFGM22, DBLP:conf/siggraph/BaeWLM023}, which provides a style imitation objective to encourage the generation of natural motions. By learning to match the distribution of the reference data rather than follow a specific trajectory, AIL allows deviations from the reference data to achieve broader task objectives, enabling better task adaptations.

In recent years, hierarchical methods have shown significant progress in physics-based character animation~\cite{DBLP:journals/tog/PengGHLF22,DBLP:journals/tog/WonGH22, wang2024skillmimic}. These methods typically involve a two-stage training process. In the initial imitation stage, a low-level controller is trained to perform a wide range of skills by mimicking the reference motions~\cite{DBLP:journals/tog/PengGHLF22, DBLP:conf/siggraph/TesslerKGMCP23, DBLP:conf/siggrapha/DouCFKW23, DBLP:conf/siggraph/WangHW24,DBLP:journals/tog/WonGH22,DBLP:journals/tog/YaoSCL22,DBLP:conf/iclr/0002CMWHKX24, DBLP:journals/tog/ZhangYMGFPF23}. The low-level controller is usually modeled using a latent variable model. Then, a high-level controller learns to output appropriate latent variables to guide the low-level controller in solving downstream tasks.
However, hierarchical methods often struggle to adapt to out-of-domain scenarios, which poses challenges in environments requiring flexible and dynamic motions. \textcolor{black}{Recent work, such as MaskedMimic~\cite{DBLP:journals/tog/TesslerGNCP24}, explores learning policies through masked conditioning and distillation from motion tracking controllers. However, it is still trained within reference-conditioned motion tracking settings, where training conditions are derived from the reference data itself. In contrast, our framework explicitly constructs more general goal-conditioned settings and jointly trains adaptation beyond the reference data distribution while maintaining motion tracking during training.}

\textit{Character-Scene Interaction:} Generating natural interactions between characters and their environments is important yet challenging. Early works use motion graphs to sequence reference clips to produce desired interactions~\citep{DBLP:journals/tog/LeeCRHP02,DBLP:journals/tog/LeeCL06}.
However, these methods were limited to the scenarios originally captured. 
Recent works apply deep neural networks for human-scene interaction~\cite{DBLP:journals/tog/StarkeZKS19, DBLP:conf/cvpr/WangXXL021,DBLP:conf/iccv/HassanCVSYZB21}, with diffusion models enabling text-conditioned generation~\cite{DBLP:conf/eccv/YiTBPR24, DBLP:conf/siggrapha/JiangHWL0H024}. However, these kinematic-based methods can not ensure physical plausibility for the generated interactions. Physics-based control methods offer more natural interaction via dynamic simulation. Chao et al.~\shortcite{DBLP:conf/aaai/ChaoYC021} trained a repertoire of controllers from reference motion data to perform tasks such as sitting on a chair. Yu et al.~\shortcite{DBLP:journals/tog/YuPL21} proposed to train separate controllers to physically reconstruct parkour movements from videos. Adversarial imitation learning has been employed to synthesize more natural interactions within indoor scenes~\cite{DBLP:conf/siggraph/HassanGWBFP23, DBLP:conf/iclr/XiaoW0CZ0LP24, DBLP:journals/corr/abs-2503-19901}. 

Recent works have explored various robots performing terrain traversal tasks~\cite{DBLP:journals/cgf/XieLKP20, DBLP:journals/tog/Panne16, DBLP:journals/scirobotics/HoellerRSH24, DBLP:conf/corl/ZhuangFWASF023, DBLP:conf/icra/ChengSAP24, DBLP:journals/corr/abs-2406-10759, xu2025parc}, primarily emphasizing navigability across obstacles. In contrast, our approach develops a single unified controller capable of performing a wide range of visually striking parkour stunts, enabling highly diverse and dynamic interactions between the character and the environment.

\section{Overview}

This paper introduces a hybrid imitation learning (HIL) framework aimed at enhancing the realism and versatility of virtual characters across novel environments and conditions. With HIL, we train a unified controller capable of executing diverse athletic behaviors while adapting seamlessly to unseen scenarios. The motion tracking mode ensures that the controller can accurately reproduce reference motions, preserving the naturalness of the reference. Meanwhile, the adversarial imitation learning mode enhances adaptability by providing the controller with the flexibility to modify and sequence these skills to tackle unseen conditions.

To facilitate effective training across these two modes, we design parallel multi-task environments. The motion tracking mode is implemented as a motion tracking task, while the adversarial imitation learning mode is implemented using Adversarial Motion Priors (AMP)~\cite{DBLP:journals/tog/PengMALK21}. However, motion tracking controllers typically require temporal phase information or future target poses as input~\cite{DBLP:journals/tog/PengALP18, DBLP:journals/tog/TesslerGNCP24}, which are not available in the adversarial imitation learning mode, where no corresponding reference motion data is available. \textcolor{black}{This mismatch in conditioning makes it difficult to train a single policy across the two modes, as the controller must rely on different inputs to achieve their respective objectives. Different from standard motion tracking frameworks, we introduce a unified condition-driven observation space, which encodes scene context and task objectives in a consistent form across both modes. In the motion tracking mode, these goal conditions (e.g., target location, facing direction, and scene geometry) constrain the set of feasible motions and implicitly indicate the controller’s progression along a reference behavior. In the adversarial imitation learning mode, the same representation provides task-relevant information for adapting skills in response to new environments. By sharing this representation, the controller can leverage a common conditioning mechanism across modes, enabling behaviors learned from reference data to transfer more effectively to more general goal-driven scenarios.}

\section{Preliminaries}
In this work, all controllers are trained using goal-conditioned reinforcement learning (GCRL), where an agent interacts with an environment in order to optimize a reward function conditioned on a task-specific goal. At each time step $t$, the agent observes the current state $s_t$ of the environment together with a goal specification $g_t$, and samples an action $a_t$ from a policy $\pi(a_t|s_t, g_t)$. Upon executing the action, the environment transitions to a new state $s_{t+1}$, following the dynamics $s_{t+1} \sim p(s_{t+1}|s_t,a_t)$, and the agent receives a reward $r_t = r(s_t,a_t,s_{t+1}, g_t)$. The objective is to learn a policy $\pi$ that maximizes the expected discounted return $J(\pi)$, defined as:
\begin{equation}
J(\pi) = \mathbb{E}_{p(\tau|\pi)} \left[ \sum_{t=0}^{T-1} \gamma^t r_t \right]
\end{equation}
where $p(\tau|\pi) = p(s_0) \prod_{t=0}^{T-1} p(s_{t+1}|s_t, a_t) \pi(a_t|s_t, g_t)$ represents
the likelihood of a trajectory $\tau = \{s_0, a_0, r_0, s_1, ..., s_{T-1}, a_{T-1}, r_{T-1}, s_T\}$ 
under the policy $\pi$.
Here, $p(s_0)$ is the initial state distribution, $T$ represents the time horizon of a trajectory, and $\gamma \in [0,1]$ is the discount factor.

\section{Hybrid Imitation Learning}
In this section, we introduce our hybrid imitation learning (HIL) framework. Our framework combines two training modes: motion tracking and adversarial imitation learning. We observed that each technique in isolation results in suboptimal behavior, such as mode collapse (using only a small subset of skills), quality degradation (unnatural behaviors), or inability to adapt to scene changes (robustness). Our hybrid framework combines the strengths of motion tracking and adversarial imitation learning, allowing the controller to both reproduce behaviors from the dataset and adapt them to new environments and task conditions.

The HIL framework is implemented through parallel multi-task environments. In the motion tracking mode, the controller is trained to track the reference motion precisely, frame by frame. In the adversarial imitation learning mode, the controller is trained with goal-conditioned tasks such as navigating obstacles or maintaining a specified heading direction. By exposing the character to goals and scenes beyond those depicted in the reference dataset, this mode encourages robustness and adaptability. By training on both tasks in parallel, the model acquires precise skills and generalizes across novel scenes and conditions.

\subsection{Motion Tracking}
In prior motion tracking systems, time phase variables~\cite{DBLP:journals/tog/PengALP18, DBLP:conf/nips/0007K20} or target poses~\cite{DBLP:conf/iccv/0002CWKX23,DBLP:journals/tog/TesslerGNCP24} are commonly used. 
However, to develop a unified controller capable of adapting to diverse scenes, we cannot rely on these inputs, as they are unavailable in novel environments where no reference data is available. Instead, a consistent observation space is crucial, as it forces the controller to adopt similar behaviors for both motion tracking and adversarial imitation. \textcolor{black}{To achieve this, we construct a goal-conditioned observation that encodes information such as scene geometry, target location, or directional cues in a unified representation. Empirically, we find that, together with the character state, this representation provides sufficient information for the controller to implicitly infer its progression within a motion clip and perform effective motion tracking without explicit pose or phase variables.}

Building on this finding, our controller takes the character state $s_t$ and the goal condition $g_t$ as input and outputs an action $a_t$ to enable the character to track a given reference motion.
The model is trained using a standard motion tracking objective, which encourages the character to minimize the difference between the state of the character and the reference motion at each timestep $t$:
\begin{align}
r^{track}_t & = w_p e^{-\alpha_p||\hat{p}_t-p_t||} + w_r e^{-\alpha_r||\hat{q}_t\ominus q_t||} + w_{v} e^{-\alpha_{v}||\hat{\dot{p}}_t-\dot{p}_t||} \nonumber \\ & + w_{\omega} e^{-\alpha_{\omega}||\hat{\dot{q}}_t-\dot{q}_t||} + w_h e^{-\alpha_h||\hat{h}_t-h_t||} + w_{e} \sum_{j}||\tau_j \dot{q}_j||,
\end{align}
where $w_{\{\cdot\}}$ and $\alpha_{\{\cdot\}}$ are weights used to balance different reward terms. This reward encourages the character to imitate the position $\hat{p}$, rotation $\hat{q}$, linear velocity $\hat{\dot{p}}$, angular velocity $\hat{\dot{q}}$, and the root height $\hat{h}$ specified by the reference motion. An energy penalty is applied to encourage smoother motion and mitigate jittering~\cite{DBLP:conf/siggraph/LeeSYWW23}. A detailed description of the reward functions is available in the Appendix. 

Incorporating motion tracking into the hybrid training framework encourages the controller to reproduce a broader range of reference behaviors. Since different obstacles are associated with different reference motions, the tracking objective encourages the policy to utilize different interaction patterns across task conditions. Sharing the same observation representation across the two modes also provides a consistent conditioning mechanism between motion tracking and adversarial imitation learning.

\subsection{Adversarial Imitation Learning}
The adversarial imitation learning mode is designed to enhance the controller's adaptability, enabling it to perform natural behaviors in novel conditions and scenes that do not exist in the reference dataset. The objective used in this mode combines a task reward and a style reward. 

The task reward is defined in a goal-conditioned manner and may vary depending on the task. For example, it can encourage the character to navigate towards a specified target or to maintain a desired heading direction. This component provides task-level guidance, which encourages the character to solve environment-specific objectives rather than simply replicating demonstrations.

The task reward is combined with a style reward derived from a discriminator $D(s_{t-n:t},c_{t-n:t})$. The goal of the discriminator is to differentiate between real data, sampled from the reference dataset, and "fake" data generated by the policy. The discriminator is provided with the $n$-step history of previous states $s_{t-n:t}$ 
and scene conditions $c_{t-n:t}$, and then predicts whether the transitions are from the reference motion dataset $\mathcal{M}$ or produced by the policy $\pi$. By adding the goal condition to the discriminator, it allows the discriminator to evaluate the naturalness of a motion, and also the motion's suitability for the current condition. The discriminator is trained using a binary classification loss \citep{DBLP:conf/nips/HoE16}, including a gradient penalty regularizer~\cite{DBLP:journals/tog/PengMALK21}:
\begin{align}
\mathop{\min}_{D} \quad &-\mathbb{E}_{d_{M}(s_{t-n:t},c_{t-n:t})} \log(D(s_{t-n:t},c_{t-n:t})) \nonumber \\ &-\mathbb{E}_{d_{\pi}(s_{t-n:t},c_{t-n:t})} \log(1 - D(s_{t-n:t},c_{t-n:t})) \nonumber \\
& +w_{gp}\mathbb{E}_{d_{M}(s_{t-n:t},c_{t-n:t})}\left\|\nabla_{\phi}D(\phi)\big|_{\phi=(s_{t-n:t},c_{t-n:t})}\right\|^{2},
\end{align} 
where $w_{gp}$ is a manually specified weight.
Similar to AMP~\cite{DBLP:journals/tog/PengMALK21}, the style reward is then given by: $r^{style}_t = -\log (1-D(s_{t-n:t},c_{t-n:t}))$, which encourages the policy to produce more natural motions while also utilizing the appropriate skills for interacting with a particular scene.

The final adversarial objective is a weighted combination of task and style rewards:
\begin{equation}
r_t=w^{task}r^{task}_t+w^{style}r^{style}_t, 
\end{equation}
with $w^{task}$ and $w^{style}$ being weights for each objective. To bridge the behavior in the two modes, a style reward $r^{style}_t$ is also added to the tracking reward $r^{track}_t$ with the same weights in the tracking mode. This adversarial objective encourages the controller to adapt and compose skills from the motion dataset as necessary to clear new obstacles and scenes. 

\begin{figure}[t]
    \centering
\includegraphics[width=1\columnwidth]{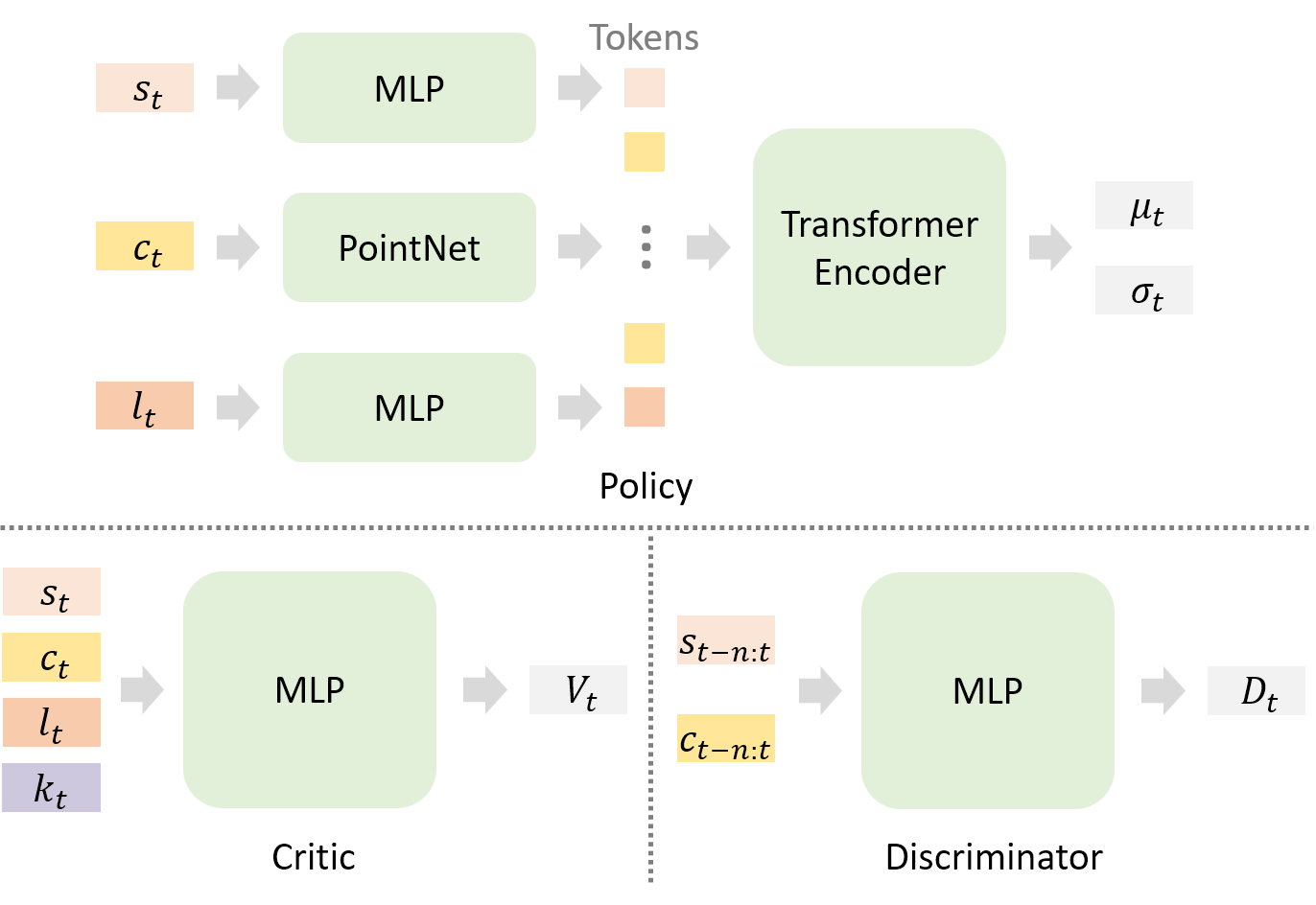}
\vspace{-0.2in}
\caption{Network architectures of the policy, the critic, and the discriminator. The policy takes character state $s_t$, scene point cloud $c_t$, and target goal location $l_t$ as input to output the action. The critic additionally takes a task indicator variable $k_t$ as input. For the discriminator, n-step state transitions $s_{t-n:t}$ and scene point cloud $c_{t-n:t}$ are provided.}
    \label{fig:architecture} 
\end{figure}

\section{Tasks}

Our goal is to develop a unified controller that can not only replicate precise athletic skills but also adapt these skills to diverse environments. To achieve this, in addition to carefully designing a multi-task environment, we incorporate several key design choices into the system. We first describe the shared design components that enable the learning across two modes, and then detail how these designs are instantiated for parkour and heading task.

\textbf{Character state}. The simulated character is constructed based on the SMPL human model~\cite{DBLP:journals/tog/LoperM0PB15}. The character's state $s_t = (h_t, p_t, q_t, \dot{p}_t, \dot{q}_t)$ is represented by a set of features that describes the configuration of the character's body: 
\begin{itemize}
    \item $h_t$: height of the root from the ground
    \item $p_t$: positions of each joint in the local coordinate frame
    \item $q_t$: rotations of each joint in the local coordinate frame
    \item $\dot{p}_t$: joint linear velocity in the local coordinate frame
    \item $\dot{q}_t$: joint angular velocity in the local coordinate frame
\end{itemize}
The root is designated as the pelvis. The character’s local coordinate frame is defined with the origin located at the root, the x-axis oriented along the root link’s facing direction, and the z-axis aligned with the global up vector. 

\textbf{Goal condition}. To construct a unified observation space effective across both training modes, we condition the policy on task-specific goal variables $g_t$. The specific form of $g_t$ differs across tasks, as detailed in the following section.

\textbf{Action}. The simulated humanoid is actuated using proportional derivative (PD) controllers. The policy  $\pi(a | s, g) = \mathcal{N}(\mu_{\pi}(a | s, g), \Sigma_{\pi})$, is modeled as a multi-dimensional Gaussian, where the mean $\mu_{\pi}$ is predicted by the model, and the covariance matrix $\Sigma_{\pi}$ is defined using manually-specified values $\sigma_{\pi} = 0.055$.

\begin{figure*}[t]
    \centering
\includegraphics[width=1.\textwidth]{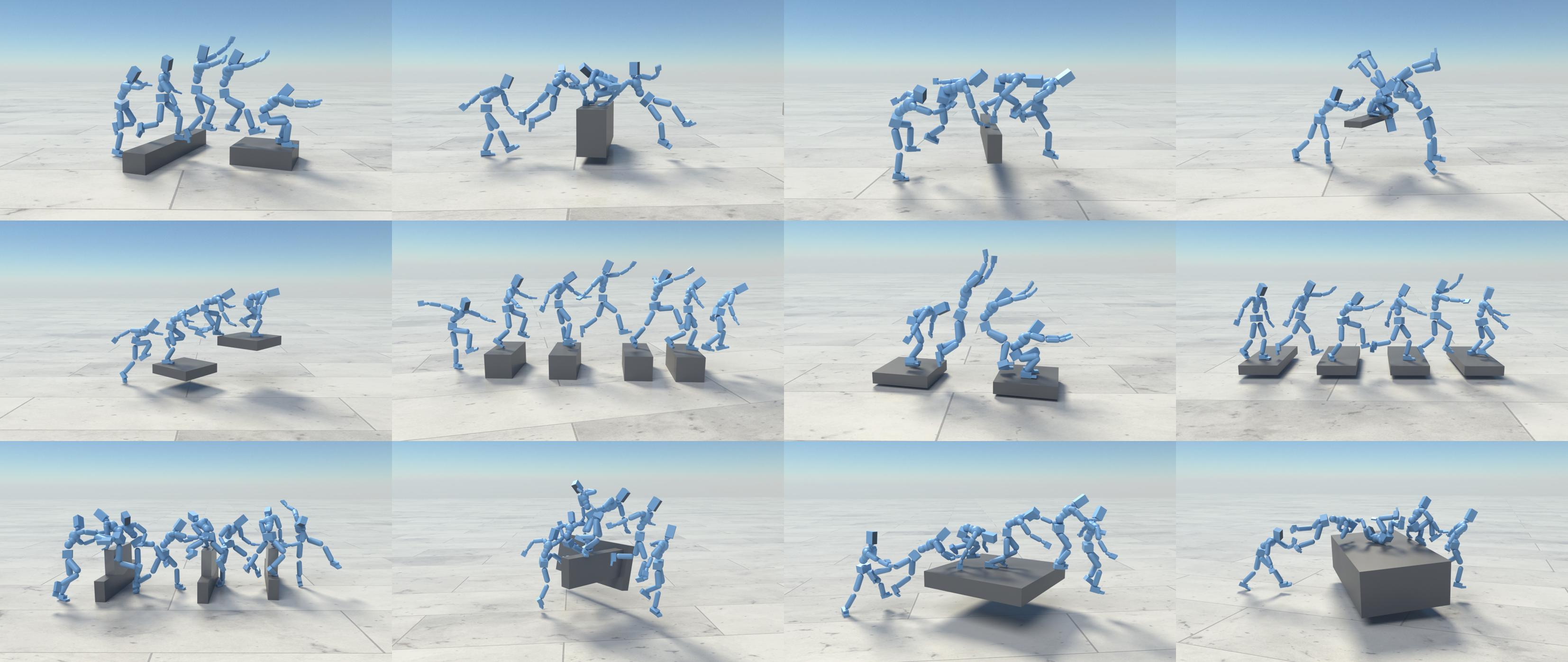}
\vspace{-0.2in}
    \caption{Our controller enables physically simulated characters to perform a wide variety of interactions.}
    \label{fig:ours}
\end{figure*}

\subsection{Parkour Task}
We first describe the task-specific design of parkour task, which focuses on agile scene interactions and obstacle traversal.

\textbf{Goal condition}. For the parkour task, the observation includes a character-centric point cloud $c_t$, defined as the closest $N$ points from the scene to the character’s root, together with a target location $l_t$. 
\textcolor{black}{In the motion tracking mode, $l_t$ is sampled from a future root position 1--2 seconds ahead along the reference trajectory. In the adversarial imitation learning mode, $l_t$ is sampled near upcoming obstacles by using the corresponding future root position and perturbing it with Gaussian noise sampled from $\mathcal{N}(0, 0.2)$.}

\textbf{Reward.} The reward encourages the root position $p^{root}_t$ to move toward the target location $l_t$:  
\begin{equation}
r^{\text{task}}_t = 
w^{prog} \Big( \|p^{root}_{t-1} - l_{t-1}\|_2 - \|p^{root}_t - l_t\|_2 \Big) 
+ w^{reach} \, r^{reach}_t,
\end{equation}
where $r^{reach}_t$ is a one-time bonus upon reaching $l_t$. This formulation promotes steady progress toward the goal and rewards successful traversal of obstacles.

\textbf{Model architecture}.
The architecture of the model for parkour task is illustrated in Figure~\ref{fig:architecture}.
The \textbf{policy} $\pi$ is modeled using a transformer-based architecture, which outputs an action distribution based on the current character state and the surrounding scene. 
PointNet~\cite{qi2017pointnet} is employed to extract features from the closest $N$ points in the scene point cloud, which are then encoded as $N$ tokens and processed by a transformer. Two multilayer perceptron (MLP) neural networks are utilized to encode character state and target goal location to the tokens for the transformer.
This architecture enables the controller to effectively integrate multi-modal observations to adapt to new and challenging scenes. 
While the policy is modeled by a transformer, the \textbf{critic} is modeled with a simple MLP. Given our hybrid training setup, with different objectives for different training modes, the critic is provided with an additional binary task indicator variable $k$ as the input. This privileged information is used solely by the critic to distinguish between the different training modes and is not provided as input to the policy. The character state, scene point cloud, target goal, and task indicator are flattened and concatenated to form the input for the critic, which then predicts the value $V$ for policy updates.
The \textbf{discriminator} is modeled by another MLP. It receives as input the flattened and concatenated features from the character state transitions and scene point cloud and outputs the discriminator logits $D$, which is used to compute the style reward.

\textbf{Dataset}. For parkour, training the controller requires reference data that contains both human motions and the corresponding scenes. Because motion capture data for parkour is scarce, we extract motion data from online videos. First, a vision-based pose estimator, TRAM~\cite{DBLP:conf/eccv/WangWLD24}, is applied to each video. However, TRAM struggles with precise ground estimation due to the complexity of parkour movements. We address this by employing the body orientation hints~\cite{DBLP:journals/tog/YuPL21}, 
which estimate the body’s up-vector in 3D space using the angle between the view up-vector and the body up-vector in the image plane. This approach refines the global orientation of the body relative to the ground. To construct the corresponding scene for simulation, we develop an interactive scene annotation tool that allows annotators to manually position basic box geometries to replicate the interaction affordances.

The kinematic motions paired with the scene annotations still exhibit several artifacts, such as human-obstacle collision, noticeable jittering, and unnatural sliding. To address these issues, we refine the collected motions using a physics-based motion tracker~\cite{DBLP:journals/tog/TesslerGNCP24}. This refinement step provides high-quality motions~\cite{DBLP:journals/tog/PengKMAL18}, which are used for computing motion tracking rewards, and sampling reliable Perturbed State Initialization (PSI). In addition, it is essential to use high-quality motions as the positive samples for the adversarial motion priors~\cite{DBLP:journals/corr/abs-2407-00187}. These refined motions ensure physically accurate interactions, facilitating the training process.

We collect 19 reference motion clips with obstacles from YouTube, totaling 30 seconds across 15 skills, each demonstrating a single skill. During training, we randomly sample five obstacles from this data to form a sequence, creating the target-following task. The tracking task involves following a single motion clip within the sequence. Obstacle placement details are provided in the appendix.

\subsection{Heading Task}
To show the generality of our framework, we also apply HIL to a heading task, which focuses on directional and orientational control.

\textbf{Goal condition}.
For the heading task, the goal condition follows the ASE formulation~\cite{DBLP:journals/tog/PengMALK21}. The controller is provided with a target heading direction $\hat{d}_t$ and a target facing direction $\hat{f}_t$, both represented as 2D unit vectors on the ground plane. These variables encourage the character to move along $\hat{d}_t$ while aligning its orientation with $\hat{f}_t$.

\textbf{Model architecture}.
The model architecture is simplified from the parkour task by removing the PointNet module and the scene point cloud $c_t$. In addition, the policy and critic inputs replace the target location $l_t$ with features of the goal variables $(\hat{d}_t, \hat{f}_t)$, representing the target heading and facing directions.

\textbf{Reward}. For the heading task, the reward encourages the velocity $v_t$ to match the target heading $\hat{d}_t$ with target speed $v^{*}$, and the facing orientation $\hat{q}_t$ to align with the target facing $\hat{f}_t$:  
\begin{equation}
r^{\text{heading}}_t =
w^{vel} \exp\!\Big(-\alpha \big(v^{*} - \hat{d}_t^\top v_t\big)^{2}\Big)
+ w^{face} \, (\hat{f}_t^\top \hat{q}_t),
\end{equation}
where $w^{vel}$ and $w^{face}$ balance velocity and facing alignment, and $\alpha$ is a scale parameter.

\textbf{Dataset}.
\textcolor{black}{We utilize the sword-and-shield dataset from~\cite{DBLP:journals/tog/PengGHLF22}, which contains approximately 7 minutes of motion capture data and is significantly larger and more diverse than the parkour dataset used in our other experiments.} These motions naturally capture behaviors such as advancing toward an opponent, retreating, turning, and maintaining orientation while moving, making them well-suited for learning heading and facing control.

\subsection{Perturbed State Initialization and Early Termination}
Following prior work, reference state initialization (RSI) can significantly improve the training efficacy of motion tracking controllers~\cite{DBLP:journals/tog/PengALP18}. However, initializing the character directly from states sampled from the reference motions poses two limitations: the system can exhibit (1) limited adaptability to disturbances and (2) difficulty in transitioning between skills. For instance, if a character needs to transition from performing skill A to skill B, and the end state of skill A does not closely match the starting state for skill B, motion tracking methods may struggle to transition effectively. To promote smooth skill transitions, we train a more robust controller capable of executing skills from a wide range of initial states. To achieve this, we incorporate \textit{perturbed state initialization} (PSI) during training, 
which applies Gaussian noise to the initial states sampled from the reference motions. This technique improves the robustness of the controller to a wider range of states, while also improving the model's ability to transition between different skills. 

Empirically, PSI improves task completion in the adversarial mode and mitigates mode collapse by promoting smoother skill transitions. When it is challenging to learn transitions between skills, controllers trained with reinforcement learning tend to adopt a "general`` approach, 
relying on simpler skills to perform various interactions. This dependency on a limited set of skills often leads to mode collapse. By enhancing the controller's capability to transition between actions, PSI encourages the system to utilize a broader spectrum of skills, thereby increasing skill diversity.

Early termination (ET) is another commonly used technique in motion imitation frameworks~\cite{DBLP:journals/tog/PengALP18}. For each training mode, we apply a different termination strategy, tailored to the distinct tasks in our hybrid imitation learning. In the motion tracking mode, episodes are terminated if any joint position deviates by more than 0.5 meters from the reference motion, or when the tracking is complete. In the adversarial imitation learning mode, for parkour task, an episode is terminated if the character falls down or misses the target by more than 2 meters. For the heading task, an episode is terminated if the character’s head height falls below 0.3 meters. These early termination techniques can improve the overall sample efficiency of the training process, as well as discourage the character from adopting undesirable behaviors.

\begin{figure*}[t]
    \centering
\includegraphics[width=1.\textwidth]{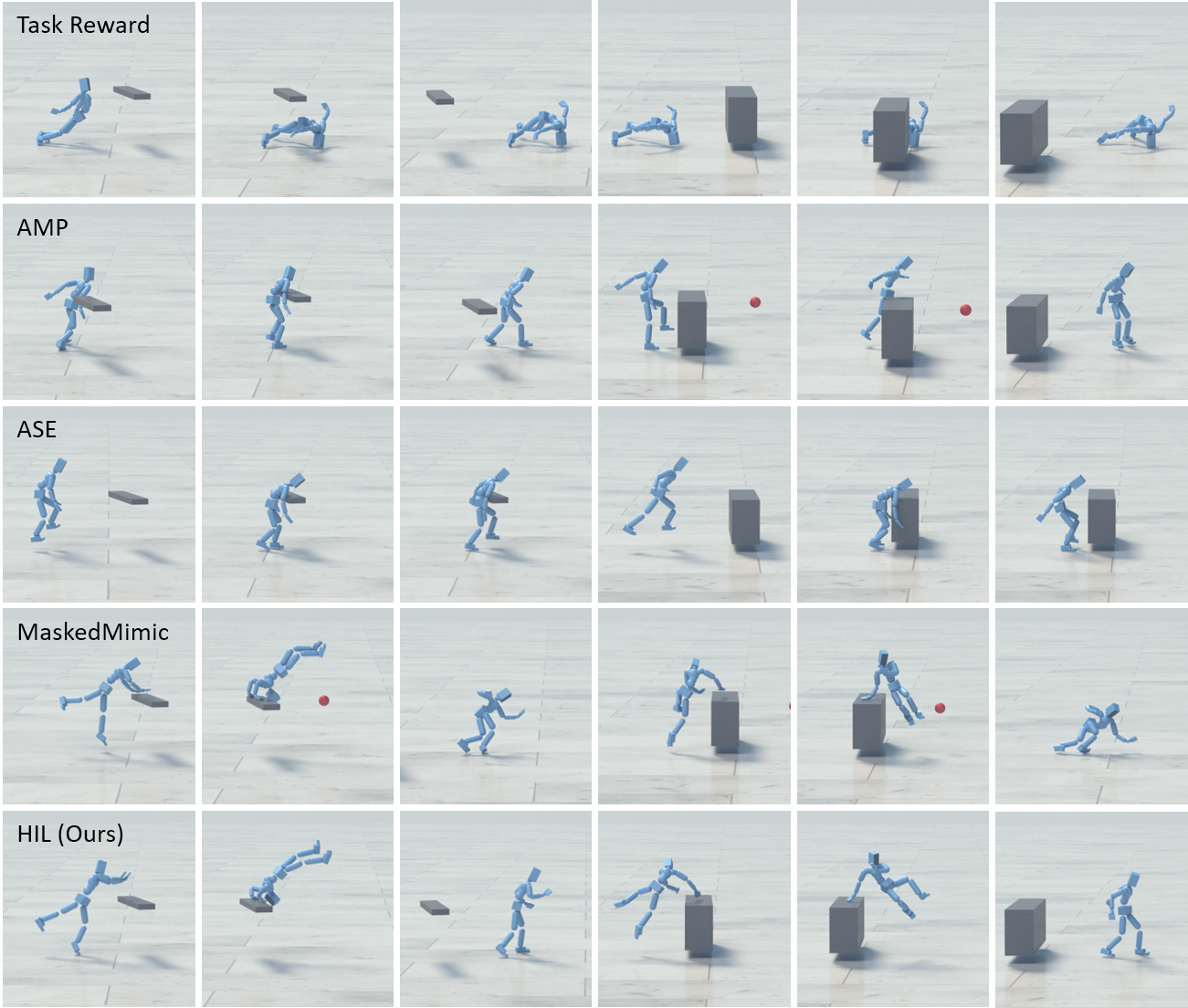}
\vspace{-0.25in}
    \caption{Qualitative comparison of different methods on the parkour task. The Task Reward, which is trained solely to optimize the task objective, tends to develop unnatural behaviors. AMP often tries to bypass obstacles, while ASE typically gets stuck in front of obstacles. MaskedMimic often falls after the first interaction. In contrast, our HIL controller is able to produce diverse and natural motions that effectively traverse long sequences of obstacles.}
    \vspace{-0.0in}
    \label{fig:compare_new}
\end{figure*}

\section{Experimental setup}
To evaluate the effectiveness of our hybrid imitation learning framework, we apply HIL to train controllers using a dataset of diverse parkour motions. We compare our method against several baselines to assess performance across a variety of new scenarios. Qualitative results are best viewed in the supplementary video.

We utilize Isaac Gym to simulate all the environments~\cite{DBLP:conf/nips/MakoviychukWGLS21}. All the experiments are trained on 4 NVIDIA V100 with a simulation frequency of 120Hz. Policies operate at 30Hz and are implemented using PyTorch~\cite{paszke2019pytorch} and optimized with Proximal Policy Optimization (PPO)~\cite{DBLP:journals/corr/SchulmanWDRK17}, using generalized advantage estimator (GAE)~\cite{DBLP:journals/corr/SchulmanMLJA15}. Initially, the model is trained on motion tracking with 4 billion samples, and then we train the model on both modes simultaneously with two billion samples, where half the environments run motion tracking and the other half adversarial imitation learning. This strategy allows the controller to first master individual skills and then adapt them to diverse scenarios. 

\begin{figure*}[t]
    \centering
\includegraphics[width=1.0\textwidth]{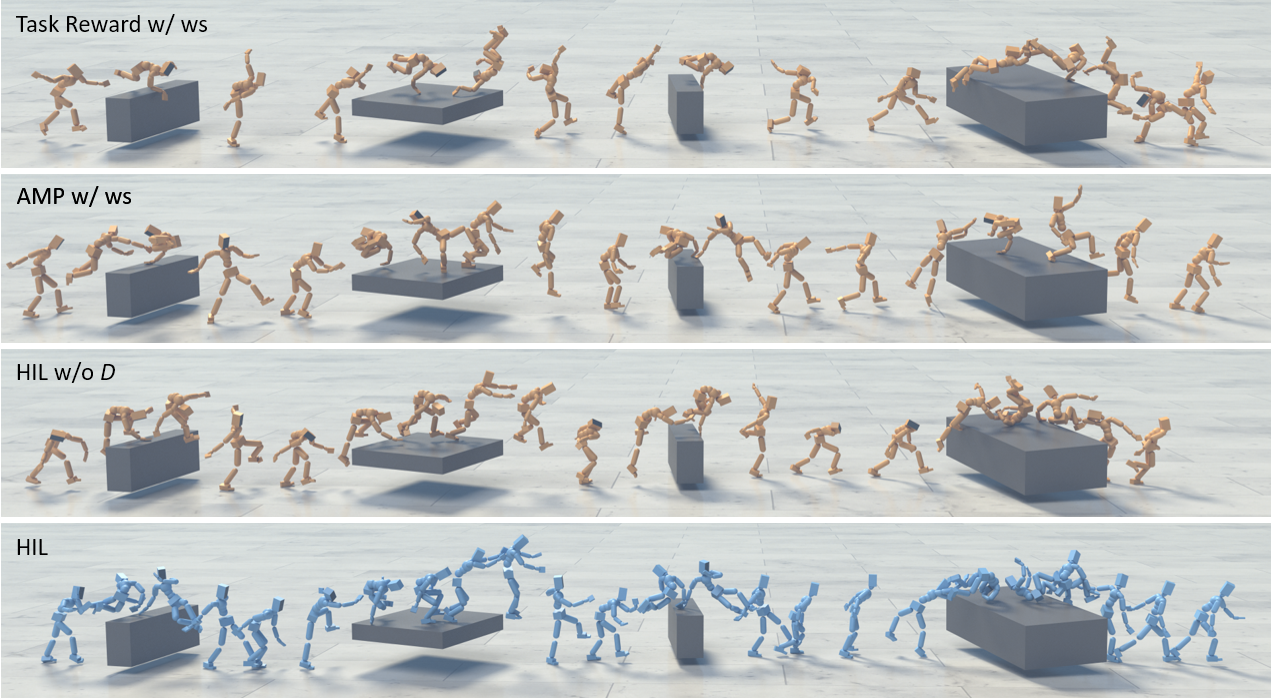}
\vspace{-0.2in}
    \caption{Motion comparisons with baselines. In this example, Task Reward w/ ws produces unnatural behaviors to clear obstacles as quickly as possible. HIL w/o D struggles to perform appropriate skills for specific obstacles, due to the independent optimization of the two tasks. AMP w/o ws suffers from severe mode collapse, repeatedly using the same skills across various obstacles. HIL generates more natural and context-aware behaviors with diverse skills.}
    \label{fig:compare}
\end{figure*}

\begin{figure*}[t]
    \centering
\includegraphics[width=1\textwidth]{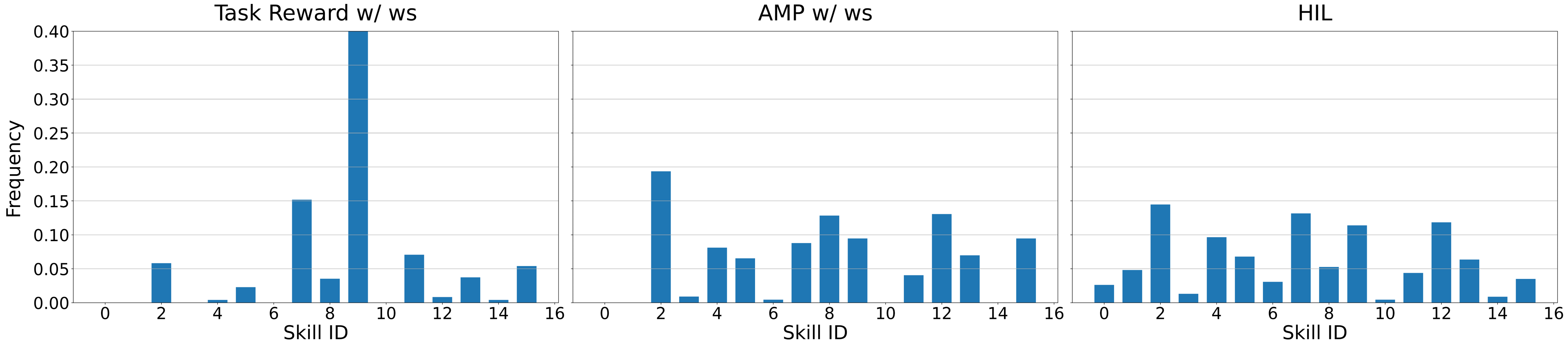}
\vspace{-0.3in}
\caption{Skill coverage comparison. The plots show the frequency of skill usage across `Task reward', AMP, and our method (HIL). `Task reward' exhibits significant bias, over-relying on certain skills, while AMP also suffers from mode collapse, with less skill diversity. HIL demonstrates broader skill usage, effectively utilizing a diverse range of skills and achieving more balanced coverage of the reference dataset.}
    \label{fig:coverage}
\end{figure*}

\subsection{Baselines}
We compare our hybrid imitation learning framework against several baselines that are commonly used in physics-based character control. We have also explained the scene representation selection in the Appendix.

\emph{Task Reward.} This baseline is trained using PPO with only the task reward and no reference motion data. It focuses purely on maximizing task performance and can succeed at obstacle traversal, but often produces unnatural and unrealistic behaviors.

\emph{Task Reward w/ Warm Start.}
\textcolor{black}{Our framework shows that motion tracking can be trained by conditioning on task-specific goals (such as scene geometry, target location, or heading/facing direction). This baseline is initialized from the same motion tracking policy used in HIL and then further trained using only the task reward.}

\emph{AMP} ~\cite{DBLP:journals/tog/PengMALK21} employs a discriminator to provide a style reward, encouraging natural human-like behavior. However, since AMP does not incorporate motion tracking during training, it often suffers from mode collapse, repeatedly using the same skill across different scenarios.

\emph{AMP w/ Warm Start.}
\textcolor{black}{This baseline is initialized from the same motion tracking policy, and then trained using adversarial imitation learning without the tracking objective.} It can be seen as an ablation of HIL without the tracking mode.

\emph{ASE}~\cite{DBLP:journals/tog/PengGHLF22} learns reusable skill embeddings from unstructured motion data by combining adversarial imitation learning and unsupervised reinforcement learning. However, it is not designed for scene-conditioned control and struggles with tasks that require explicit interaction with various environments.

\emph{MaskedMimic}~\cite{DBLP:journals/tog/TesslerGNCP24} is a CVAE-based distillation framework that trains a student policy from a teacher policy via motion tracking task. \textcolor{black}{Since both teacher and student policies are trained only through motion tracking, the method may struggle to adapt to goal conditions out of reference data and unseen transitions between skills.}

\subsection{Evaluation Metrics} 
\label{sec:metrics}
We evaluate the performance of different methods based on naturalness and task performance. 

\textbf{Parkour task.} To assess naturalness, we examine the skill coverage and motion quality, when compared to the reference data. We evaluate whether the controller performs the correct skills for interacting with each particular obstacle, referred to as \textit{skill accuracy}, and measure the deviation from reference motions using \textit{tracking error}. Since the character aims to traverse a sequence of obstacles, the generated motions may not be synchronized with the reference motions, making frame-by-frame comparison impractical. We use Dynamic Time Warping (DTW) to align motions \cite{DBLP:books/daglib/0019158}. To estimate the \textit{skill accuracy}, we compute the DTW distance between the generated motion clip and all reference motions and identify the most similar reference motion to infer the performed skill. If the performed skill matches the expected reference motion for the obstacle, it is considered correct. Otherwise, it is incorrect. \textit{Tracking error} is defined as the DTW distance between the generated motion and the ground-truth reference motion corresponding to the specific obstacle. Additional details on DTW computation are provided in the appendix.

We also evaluate the controller's effectiveness on the target following task. Task performance is quantified using the average \textit{task completion} rate, which evaluates whether the character successfully traverses a random sequence of five obstacles and reaches the endpoint. To test the controller's robustness to scene variations, Gaussian noise $\mathcal{N}(0,0.03)$ is applied to perturb the position and orientation of the obstacles, and Gaussian noise $\mathcal{N}(1,0.03)$ is used to perturb the obstacle's scale.

\textbf{Heading task.}
For heading control, we report two error metrics: direction error and facing error. The \emph{direction score} is defined as the cosine similarity between the normalized root velocity vector $v_t$ and the target heading direction $\hat{d}_t$,
and the \emph{facing score} is defined as the cosine similarity between the character’s facing vector $\hat{q}_t$ and the target facing direction $\hat{f}_t$. 
Both metrics are averaged over evaluation episodes, with higher values indicating more accurate control. We also measure average evaluation return, which reflects the cumulative reward achieved by the policy across test episodes.

\begin{table}[t]
\caption{\textcolor{black}{Quantitative comparison of our method and baselines under noisy, unseen scene variations. Obstacle position, orientation, and scale are perturbed during evaluation. HIL achieves the best skill accuracy and tracking error while maintaining competitive task completion, demonstrating that it can adapt reference skills to unseen obstacle configurations.}}

\vspace{-0.1in}
\label{tab:table1}
\begin{minipage}{\columnwidth}
\small
\begin{center}
\resizebox{\linewidth}{!}{%
\begin{tabular}{|l|ccc|}
  \hline
  \bf{Method} & \bf{Skill Accuracy$\uparrow$} & \bf{Track Error$\downarrow$} & \bf{Task completion$\uparrow$} \\ 
  \hline
  Task Reward & 0.00 & 1.82 & 0.81 \\
  AMP  & 0.06 & 1.49 & 0.11 \\
  ASE & 0.03 & 1.63  & 0.00 \\
  MaskedMimic & 0.50 & 0.41  & 0.00 \\
  Task Reward w/ ws & 0.15 & 0.54  &  \textbf{0.86}  \\
  AMP w/ ws & 0.54 & 0.37 & 0.85  \\
  HIL (Ours) & \textbf{0.66} & \textbf{0.31} & 0.74  \\
  \hline
\end{tabular}}
\end{center}
\end{minipage}
\end{table}%

\begin{table}[t]
\caption{Ablation study results. Removing the discriminator (w/o $D$), Perturbed State Initialization (w/o PSI), or scene information in the discriminator ($D^{\text{w/o scene info}}$) significantly impacts skill accuracy, tracking error, and task completion, demonstrating the importance of these components.}
\vspace{-0.1in}
\label{tab:table2}
\begin{minipage}{\columnwidth}
\small
\begin{center}
\resizebox{\linewidth}{!}{%
\begin{tabular}{|l|ccc|}
  \hline
  \bf{Method}  & \bf{Skill Accuracy$\uparrow$} & \bf{Track Error$\downarrow$} & \bf{Task completion$\uparrow$} \\ \hline
  
  w/o $D$ & 0.53 & 0.36  & 0.62  \\
  w/o PSI & 0.5 & 0.37 & 0.52  \\
  $D^{\text{w/o scene info}}$ & 0.38 & 0.39 & \textbf{0.75} \\
  w/o $k$ & 0.52 & 0.40 & 0.73 \\
  HIL (Ours) & \textbf{0.66}  & \textbf{0.31} & 0.74 \\
  \hline
\end{tabular}}
\end{center}
\end{minipage}
\end{table}%

\section{Parkour Results}
To evaluate the effectiveness of our hybrid imitation learning method, we conduct quantitative comparisons against six baseline methods from prior work. Table~\ref{tab:table1} reports quantitative comparisons across baselines, and Figure~\ref{fig:compare_new} and Figure~\ref{fig:compare} provides qualitative examples of the learned behaviors. Our hybrid imitation learning (HIL) achieves the best balance between naturalness and task performance, with the highest skill accuracy (0.66) and lowest tracking error (0.31), while maintaining a strong task completion rate (0.74). \textcolor{black}{As described in Sec.~\ref{sec:metrics}, these results are evaluated on procedurally generated obstacle sequences with perturbations to obstacle position, orientation, and scale. This setting demonstrates that HIL can learn natural behaviors from reference motions and adapt them to new scene configurations beyond those observed in the reference data. However, when the perturbations become sufficiently large and lead to obstacle configurations that differ significantly from those encountered during training, failures can still occur.}

The Task Reward baseline, which relies purely on task optimization, can achieve very high completion rates but relies on extremely unnatural strategies. Without a warm start, the agent often collapses into degenerate behaviors such as lying on the ground and “crawling” past obstacles. With warm start from our conditional motion tracking policy, Task Reward w/ ws displays more coordinated movements and is able to run across obstacles, but the interactions remain unnatural and ignore the affordances of the scene, as Figure~\ref{fig:compare} shows. \textcolor{black}{These behaviors are partly due to limitations of the simulated humanoid and environment. The simulated SMPL humanoid model is not fully physically realistic and possesses unrealistically strong actuation capabilities, enabling motions such as excessively high jumps that are difficult for real humans. Moreover, policies optimized purely for task completion may further exploit imperfections or simplifications in the simulation environment to maximize success rates. As a result, these behaviors can achieve high task performance while sacrificing realistic motion quality.} 

The baseline AMP, trained with a task and a style reward, frequently fails to complete tasks, with a completion rate of only 0.11. The character tends to walk toward obstacles, stall for long periods, and eventually attempt to bypass them by walking around rather than clearing them. AMP w/ warm start is significantly better than AMP (0.85 completion rate), and the character is able to produce visually natural motions. However, the policy relies on a very narrow set of skills, often repeating the same vault motion regardless of obstacle type, which indicates severe mode collapse.

The ASE baseline is evaluated in its pretrained setting. Since it relies solely on a discriminator reward without any task guidance, it fails to produce meaningful interactions with obstacles and typically stalls in front of them, yielding zero task completion. This contrasts with ASE’s success in flat-ground locomotion tasks, showing that discriminator-only objectives cannot provide sufficient guidance in these dynamic parkour settings.

The MaskedMimic baseline also achieves zero completion. Both its teacher and student policies are trained only on motion tracking, where conditions are sampled from reference sequences. When presented with new conditions, such as transitioning from one interaction to another, the character fails to adapt, often falling after completing the first obstacle. This inability to generalize to new environments reflects its limitations of training purely through tracking without any explicit tasks and rewards. \textcolor{black}{We note that a hypothetical additional third-stage finetuning procedure for MaskedMimic could potentially improve adaptation. However, such a setting is conceptually similar to our Warm Start baselines. Both the MaskedMimic student policy and our goal-conditioned tracking policy are pretrained on the reference distribution and can perform well within reference-conditioned settings. Our Warm Start baselines further show that simply finetuning such pretrained policies in more general task conditions is insufficient for robust adaptation beyond the reference distribution, often leading to unnatural behaviors or reduced skill diversity. In contrast, our framework maintains motion tracking during adaptation, allowing the tracking objective to serve as a regularization signal throughout training.}

In contrast, our HIL framework integrates motion tracking with adversarial imitation, producing policies that both respect the reference motions and adapt flexibly to novel scenes. As shown in Figure~\ref{fig:compare_new} and Figure~\ref{fig:compare}, HIL learns to clear different obstacles using diverse skills. A key factor behind this success is our condition-phase observation design, which allows us to train motion tracking policies directly without relying on target poses. This design already improves the performance of baselines such as vanilla AMP and vanilla Task Reward by providing a better warm start. However, the full strength of our approach lies in its ability to jointly train motion tracking and adversarial imitation within a unified observation space. This joint training maximizes the use of reference data, enabling the controller to faithfully reproduce motions and to adapt them in novel scenarios. As a result, our method achieves the best balance between skill fidelity and adaptability in challenging parkour environments.

To further evaluate skill coverage, we analyze the frequency of each skill's occurrence. We compare our method with Task Reward w/ ws and AMP w/ ws, since they can produce meaningful interactions. We sample obstacles evenly from the reference motions and sequence them into scenes. To determine which skill a character is performing over a span of time, we use the DTW distance to find the most similar behavior in the reference motion dataset. The distribution of skills for different methods is visualized in Figure~\ref{fig:coverage}. Both Task reward and AMP suffer from mode collapse, relying on a limited set of behaviors to clear obstacles. For example, AMP frequently uses the same vault motion (Figure~\ref{fig:compare}), regardless of the obstacle's characteristics. In contrast, our method demonstrates more diverse skill usage, adapting different parkour skills appropriately for different obstacles.

\begin{figure}[t]
    \centering
\includegraphics[width=1\columnwidth]{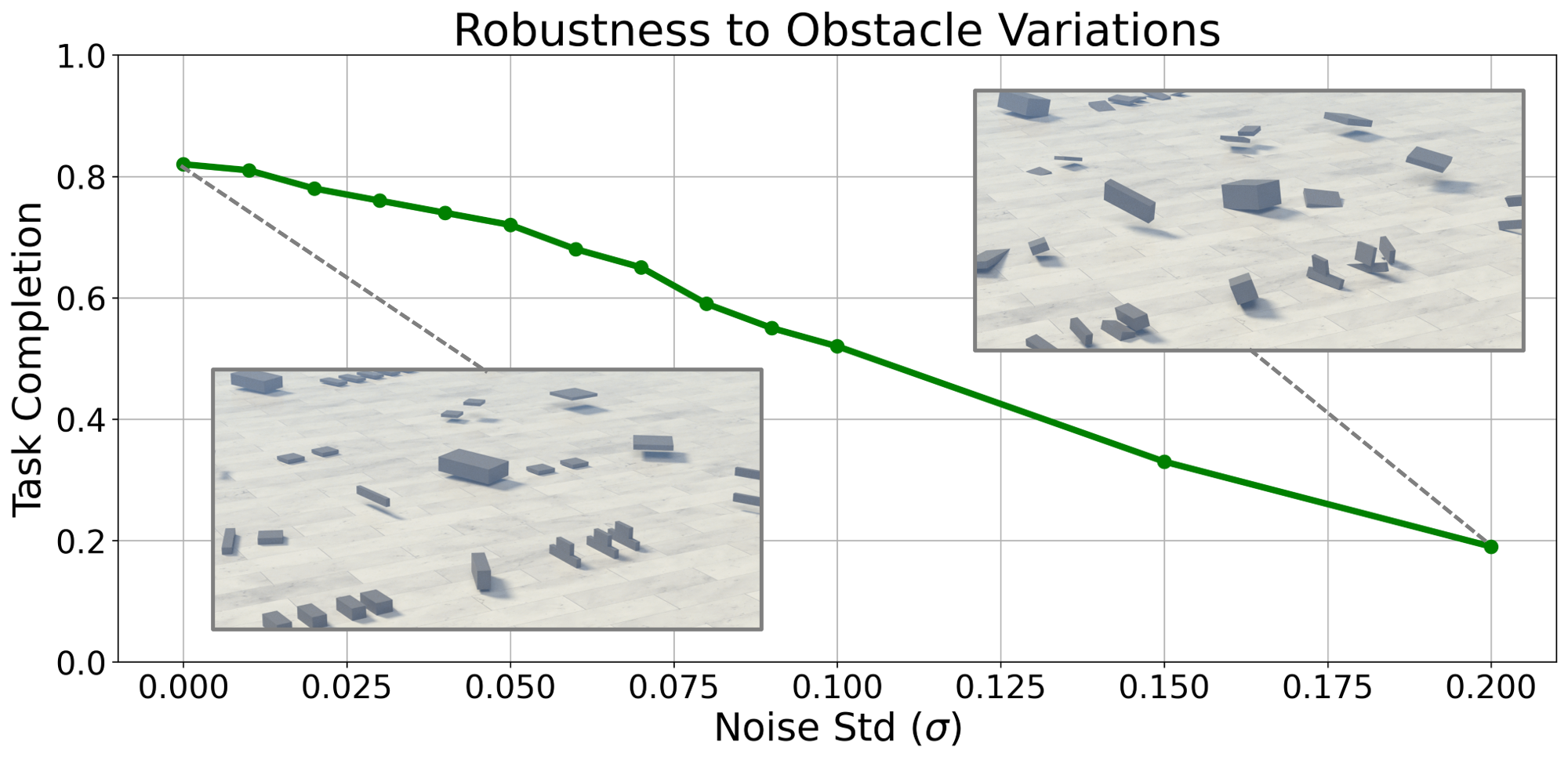}
\vspace{-0.3in}
\caption{Task completion performance as the noise variance increases, demonstrating the controller's robustness under varying levels of scene difficulty. Gaussian noise $\mathcal{N}(0,\sigma)$ is applied to perturb the position and orientation of the obstacles and Gaussian noise $\mathcal{N}(1,\sigma)$ is used to modify the obstacle scales. Two examples of obstacle courses are illustrated for noise levels $\sigma=0$ and $\sigma=0.2$}
    \label{fig:robust}
    \vspace{-0.2in}
\end{figure}

\begin{figure}[t]
    \centering
\includegraphics[width=1\columnwidth]{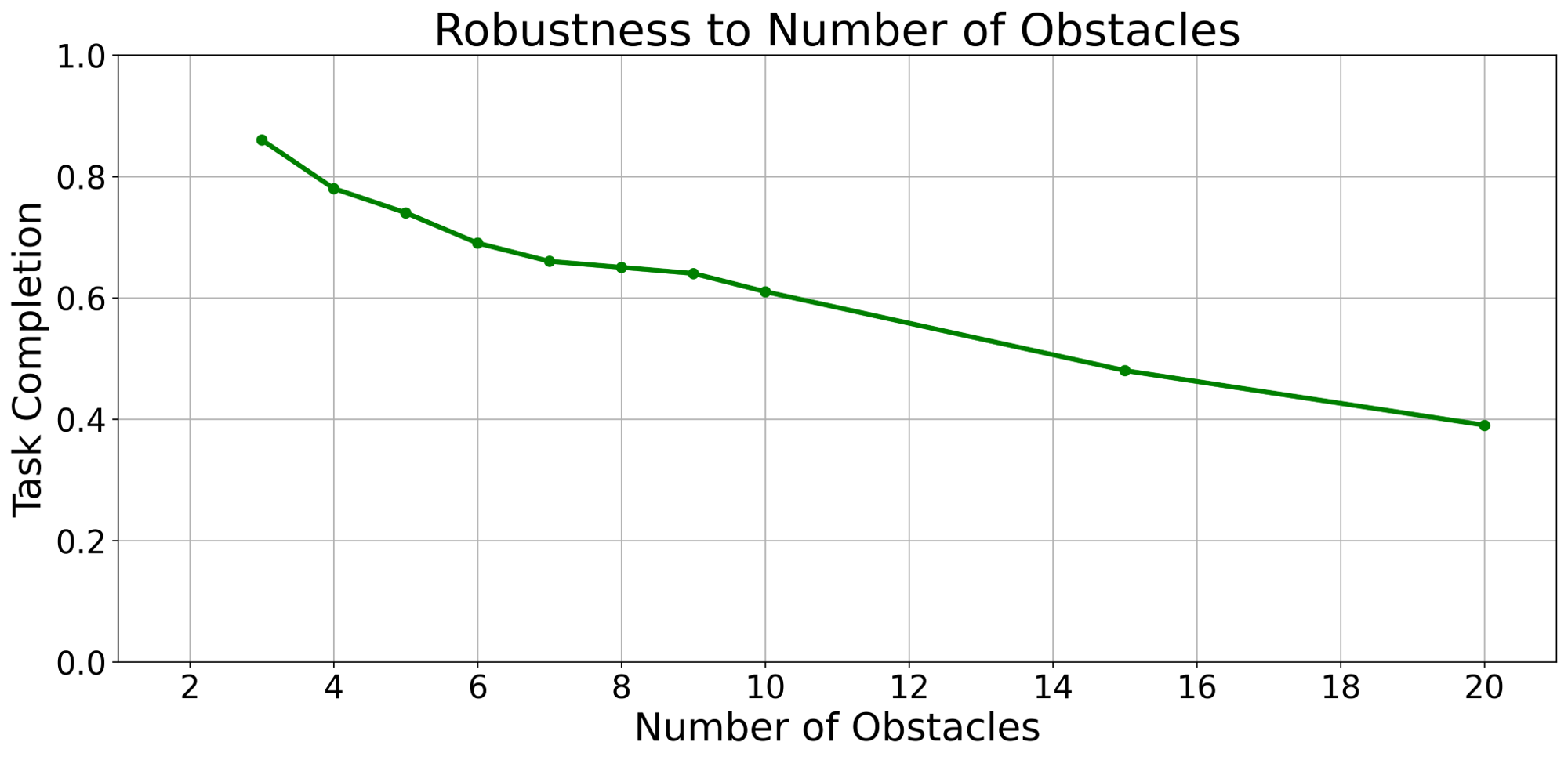}
\vspace{-0.3in}
\caption{Task completion performance as the number of obstacles increases, demonstrating the controller's robustness to clear sequential obstacles. Noise with $\sigma=0.03$ is applied to all obstacles.}
    \label{fig:robust2}
\vspace{-0.15in}
\end{figure}

\begin{figure*}[t]
    \centering
\includegraphics[width=1\textwidth]{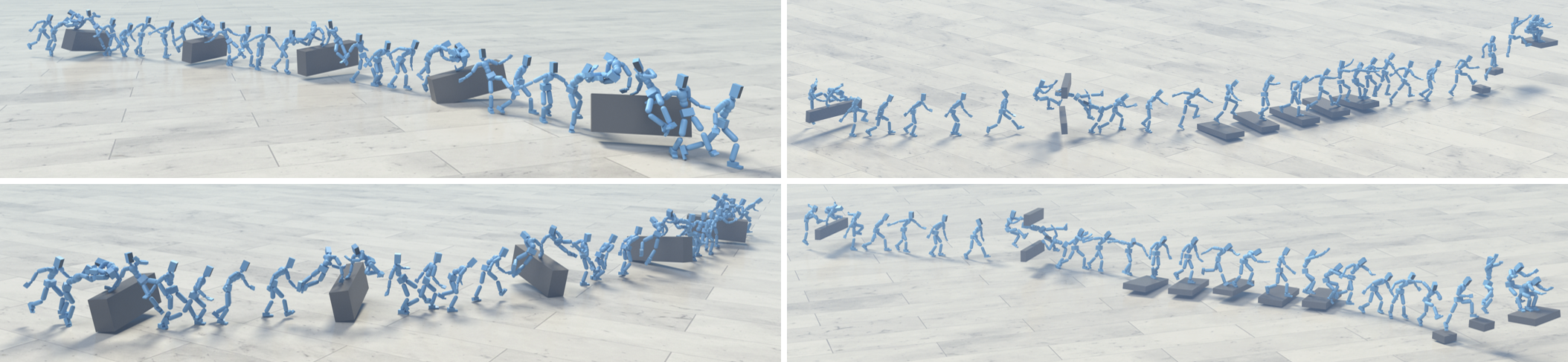}
\vspace{-0.2in}
\caption{The controller trained with HIL demonstrates remarkable robustness, allowing the character to adapt to various obstacle variations. In this example, Gaussian noise $\mathcal{N}(0,0.03)$ is applied to the position and orientation of the obstacles, and Gaussian noise 
$\mathcal{N}(1,0.03)$ is applied to the scale of the obstacles.} 
\label{fig:adapt}
\end{figure*}

\begin{figure*}[t]
    \centering
\includegraphics[width=1\textwidth]{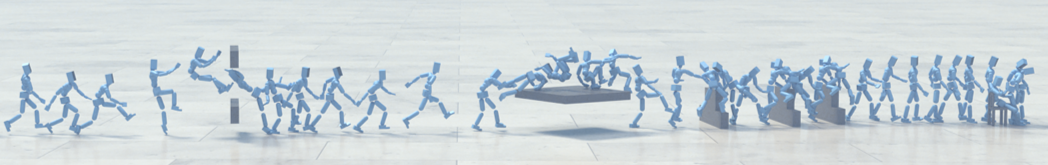}
\vspace{-0.2in}
\caption{The character performs diverse parkour skills and finishes by sitting on the chair.}
    \label{fig:generalize}
\end{figure*}

\subsection{Ablations}
We conduct ablation experiments to evaluate key components of our framework and report the results in Table~\ref{tab:table2}. First, we examine adversarial imitation learning by removing the discriminator objective (w/o $D$). The controller is still trained jointly on motion tracking and task training modes, but no style reward is provided. This variant performs significantly worse in both skill accuracy and task completion. The discriminator encourages appropriate skill selection for each obstacle and provides a reward signal that facilitates multi-task learning. By promoting motions that resemble reference trajectories, the discriminator creates synergy between motion tracking and adversarial imitation, allowing training in one mode to benefit the other. Without it, the two tasks are optimized more independently, requiring longer training for task completion. Although the controller can still track reference motions in isolation, it struggles to produce natural and effective behaviors when clearing sequences of obstacles or adapting to unseen ones.

We also ablate the Perturbed State Initialization (PSI), which adds perturbations to the character's initial state when sampling from reference motions. Disabling PSI leads to a substantial drop across all evaluation metrics. PSI enhances the policy's robustness by training the model to handle perturbations. Even when the initial state deviates from the reference, the policy can still generate behaviors that are similar to the reference over time. This robustness is particularly helpful for composing sequential skills. In our task, the state at the end of one interaction may differ from the initial state required for the next, making transitions difficult to learn. Adding perturbations improves robustness to such deviations, making skill transitions easier. Additionally, the increased robustness helps mitigate mode collapse. When transitions between skills are difficult, the RL policy can be easily optimized to execute simpler skills, neglecting the complex but necessary transitions. By making transitions easier to learn, PSI promotes a more diverse set of skill usage. In our work, PSI can be seen as a bridge between motion tracking and task training, making the optimization process more seamless and efficient.

We validate the effectiveness of scene information in the discriminator by removing it from the discriminator ($D^{\text{w/o scene info}}$). While task completion remains similar, skill accuracy and tracking error degrade. Without scene context, the discriminator evaluates state transitions in isolation, allowing behaviors that appear natural in isolation to receive high rewards, regardless of their fit with the scene. Consequently, the model may perform inappropriate skills, resulting in unnatural interactions that are misaligned with the scene. Finally, we ablate the task indicator in the critic (w/o $k$). The task indicator in the critic is also crucial, as the two modes have different reward structures (tracking vs. adversarial imitation learning). Without the task indicator, the critic fails to estimate values accurately, resulting in worse performance. We observe that the critic's loss is 5x larger when the task indicator is removed. 

\subsection{Diversity, Robustness and Generality}

In this section, we provide additional results highlighting the diversity, robustness, and generalization of our model. Figure~\ref{fig:ours} demonstrates the controller's versatility, showing interactions with different obstacles using a diverse set of parkour skills. Figure~\ref{fig:adapt} illustrates the controller's adaptability and robustness, showing how it adjusts the same skill to accommodate variations in obstacle size, position, and orientation.

To further assess the robustness of the controller, we conduct experiments where noise is added to obstacle characteristics in Figure~\ref{fig:robust}. Specifically, Gaussian noise $\mathcal{N}(0,\sigma)$ is applied to perturb the position and orientation of the obstacles, while $\mathcal{N}(1,\sigma)$ is used to modify their scale. Our model achieves a task completion rate exceeding 70\% when $\sigma=0.05$, and maintains over 50\% completion even with $\sigma=0.1$. To illustrate the effects of noise, we provide two examples of the obstacle courses with $\sigma=0.0$ and $\sigma=0.2$, which demonstrate how the obstacle characteristics are perturbed under different noise levels. We also test the controller’s generalization to longer sequences in Figure~\ref{fig:robust2}. Although trained on sequences of five obstacles with $\sigma=0$, our model achieves a 40\% task completion rate on sequences of twenty obstacles with $\sigma=0.03$. These results highlight the robustness and adaptability of our method in more challenging and noisy environments.

To demonstrate the generality of our method beyond parkour skills, we combine the parkour motion data with sitting motions from the SAMP dataset~\cite{DBLP:conf/iccv/HassanCVSYZB21}. \textcolor{black}{This introduces additional behaviors that differ significantly from parkour. We train a policy capable of seamlessly performing both dynamic parkour stunts and everyday interactions, such as sitting on chairs, as shown in Figure~\ref{fig:generalize}. Notably, the chair interaction also involves more complex object geometry compared to the simple obstacle structures used in the parkour task. These behaviors can be best viewed in the supplementary video. Together, these results suggest that our framework can incorporate diverse data and interaction types beyond a single motion domain or simplified environments, enabling a wide spectrum of behaviors ranging from highly dynamic maneuvers to everyday activities.}

\begin{table}[t]
\caption{Quantitative results on the heading task.}
\vspace{-0.1in}
\label{tab:table3}
\begin{minipage}{\columnwidth}
\small
\begin{center}
\resizebox{\linewidth}{!}{%
\begin{tabular}{|l|ccc|}
  \hline
  \bf{Method} & \bf{Direction Score$\uparrow$} & \bf{Facing Score$\uparrow$} & \bf{Avg Eval Return$\uparrow$} \\ 
  \hline
  AMP  & \textbf{0.95} & 0.94 & \textbf{266} \\
  ASE & 0.54 & 0.78  & 147 \\
  MaskedMimic & 0.79 & 0.72  & 17 \\
  HIL (Ours) & 0.94 & \textbf{0.97} & 227  \\
  \hline
\end{tabular}}
\vspace{-0.15in}
\end{center}
\end{minipage}
\end{table}%

\section{Heading Results}

\begin{figure}[t]
    \centering
\includegraphics[width=1\columnwidth]{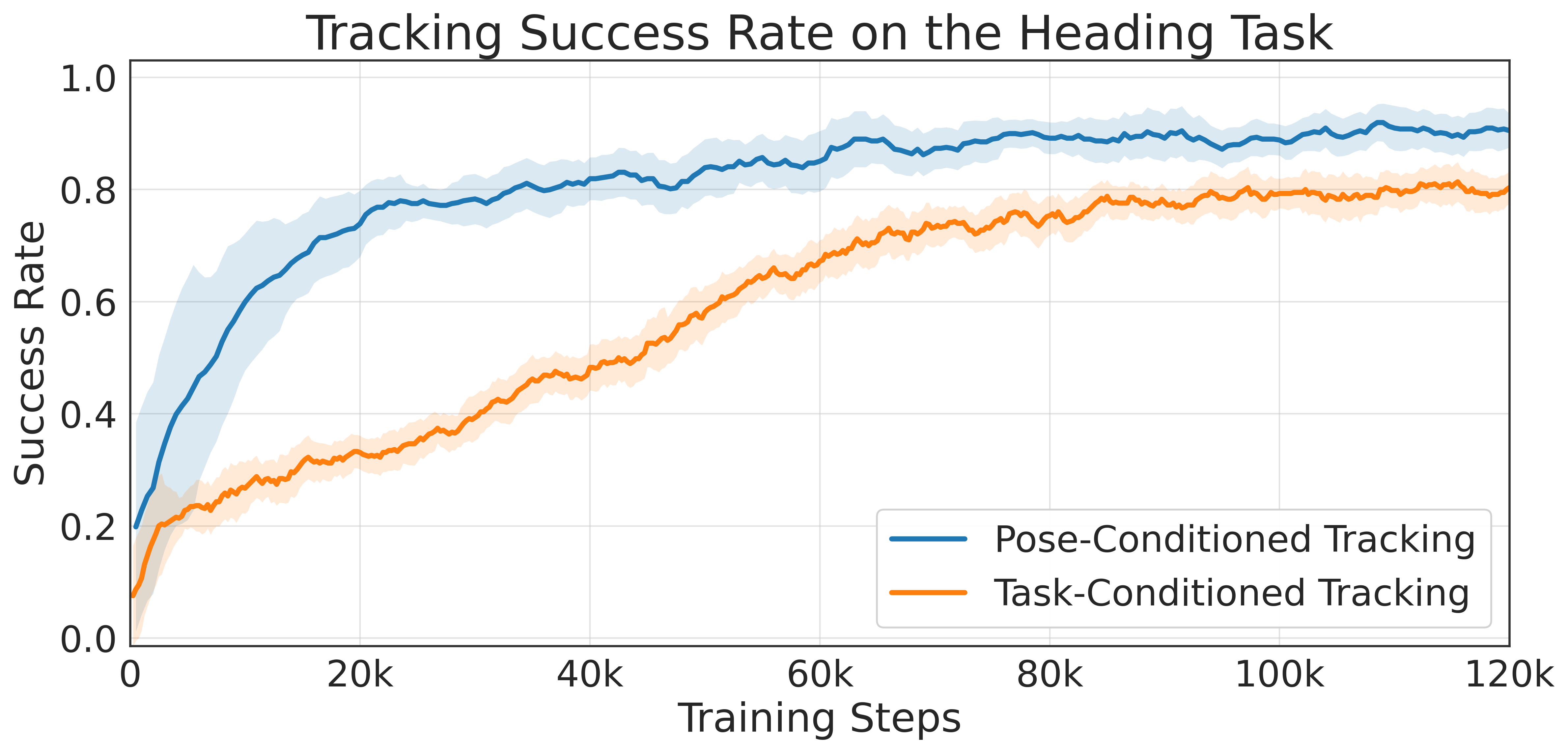}
\vspace{-0.2in}
\caption{\textcolor{black}{Tracking success rate during training on the heading task. The task-conditioned tracker achieves effective tracking behavior without explicit target poses.}}
    \label{fig:heading_tracking}
    \vspace{-0.1in}
\end{figure}

\textcolor{black}{To better understand whether task-conditioned observations can provide sufficient information for motion tracking, we conduct a controlled experiment on the heading task. We compare a standard pose-conditioned tracker~\cite{DBLP:journals/tog/TesslerGNCP24}, which receives future target pose information as input, with a task-conditioned tracker that receives only the heading and facing directions derived from the reference motion. In Figure~\ref{fig:heading_tracking}, we report the tracking success rate during training, where success is defined as whether the policy can successfully track a reference motion to its end without triggering early termination. We observe that the task-conditioned tracker is still able to learn effective tracking behavior, although it learns more slowly and converges to a slightly lower success rate than the pose-conditioned tracker. Nevertheless, the performance remains relatively close while relying only on task-level conditioning instead of explicit target poses. This result suggests that this conditioning representation can effectively support reference-guided tracking, while also naturally extending to more diverse and randomized task-condition distributions for learning general goal-conditioned controllers. Such a formulation enables a unified end-to-end framework that smoothly transitions from tracking reference behaviors to more flexible goal-driven control without requiring separate tracking-specific inputs.}

We then evaluate our HIL framework on the heading task, where the character must align both its velocity with a target heading direction and its body orientation with a target facing direction. Quantitative results are reported in Table~\ref{tab:table3}, and qualitative behaviors of our method are illustrated in Figure~\ref{fig:heading}. Additional demonstrations are provided in the supplementary video for better visual comparison.

In our evaluation, we find ASE produces more natural motions compared to AMP, as its skill embedding helps retain reference behaviors. However, ASE sometimes struggles with task performance, achieving lower returns and having lower task scores than AMP. AMP, on the other hand, achieves higher average evaluation returns, but its behaviors often appear less natural. We also observe that both ASE and AMP tend to utilize only a limited subset of behaviors from the reference dataset. The MaskedMimic baseline performs poorly across all metrics, reflecting its inability to generalize beyond reference motion tracking. Since its policies are trained solely to imitate reference motion, it fails to adapt when given new heading or facing goals, with the character quickly losing balance and falling.


\textcolor{black}{Unlike the parkour experiments, the heading task is trained with a substantially larger motion dataset containing locomotion and combat behaviors. This setting provides a different test of our framework: rather than learning from a small set of obstacle-specific skills, the controller must leverage a broader distribution of reference behaviors while adapting them to new heading and facing goals. HIL is able to capture this diversity and express multiple behavior modes, such as advancing, retreating, turning, and swinging the sword to satisfy directional goals.  These results suggest that HIL can scale beyond small curated motion sets and use larger reference datasets to produce natural, robust, and diverse goal-conditioned behaviors.}

\begin{figure}[t]
    \centering
\includegraphics[width=1\columnwidth]{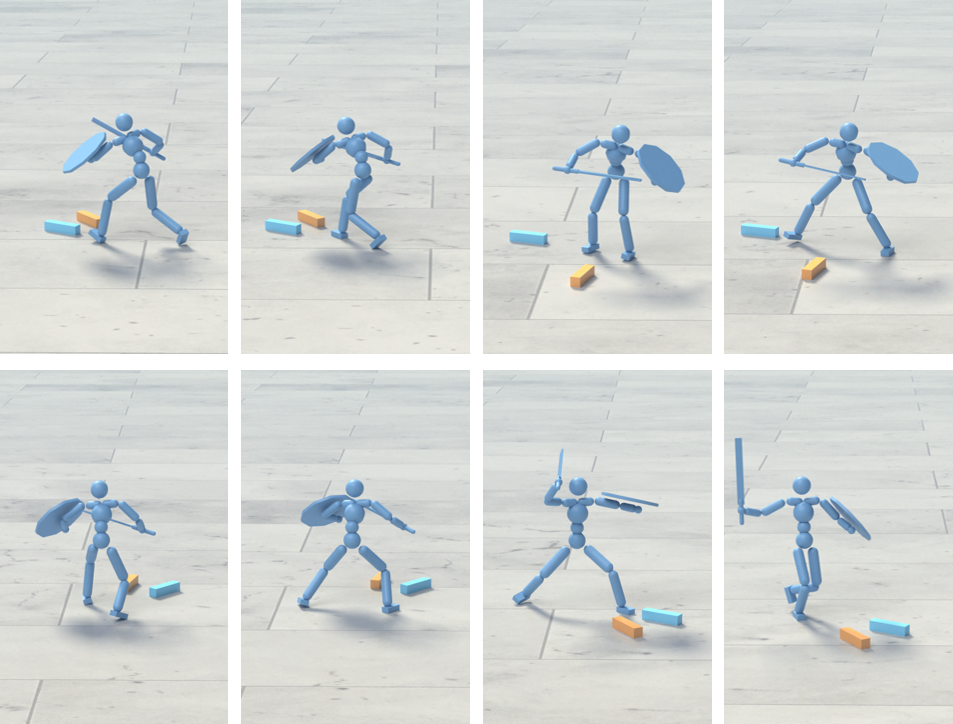}
\caption{Qualitative examples of heading task, showing that HIL produces natural and diverse behaviors.}
    \label{fig:heading}
    \vspace{-0.2in}
\end{figure}

\section{Discussion}
This work presents a simple but effective hybrid imitation learning framework for dynamic athletic control, combining motion tracking with adversarial imitation. To support efficient training, we design parallel multi-task environments and introduce a unified goal-conditioned observation space along with a perturbed state initialization strategy. Our method achieves high-quality motion generation, diverse skill usage, and competitive task performance.

\textcolor{black}{Despite these promising results, several limitations remain. While the generated motions generally exhibit high fidelity motions, occasional artifacts such as unnatural recovery behaviors when tripping over obstacles can still occur. The current parkour setup also assumes sequential obstacle courses with relatively simple box-based geometries, which limits generalization to more complex, non-linear environments and richer interaction patterns. Although our policy operates on a point cloud representation and conditions both the policy and discriminator on scene context, the diversity of interactions that can be learned is still constrained by the available reference data.}

\textcolor{black}{The heading task provides an initial indication that the proposed hybrid formulation can scale to larger motion datasets. In this setting, the controller is trained on a substantially larger motion dataset, and HIL remains effective, improving motion naturalness and behavioral diversity compared with baselines. This suggests that the framework can benefit from increased motion coverage. For parkour, however, scaling is more challenging because the data must capture not only diverse motions, but also corresponding scene geometry and interaction affordances. In our current pipeline, these motion-scene pairs are constructed manually using simplified obstacle abstractions, making large-scale data collection difficult.} \textcolor{black}{Future work could benefit from improving data diversity and interaction complexity. Recent advances in 3D reconstruction and human motion estimation could help automate the collection of paired motion-scene data from videos, reducing the need for manual scene annotation and enabling training in richer environments. Such data would make it possible to extend the framework beyond simple obstacle courses to non-sequential layouts and more complex scene geometries.} 

\textcolor{black}{While our experiments focus on parkour and heading control, the technique is not limited to these domains. More generally, our approach is applicable to goal-conditioned interaction tasks that require seamless skill composition and adaptation to novel environments. Unlike previous motion tracking formulations that rely on explicit temporal phase variables or target poses, our framework instead conditions behavior on character state, task objectives, and scene geometry within a unified observation space. For the tasks considered in this work, we find that these spatial and task constraints provide sufficient structure for coherent behavior progression while still allowing adaptation to unseen environments where reference trajectories are unavailable. This formulation could potentially extend to broader interaction-rich settings, such as indoor interaction or collaborative multi-agent scenarios, offering a path toward more general-purpose controllers for physics-based character animation in complex and interactive environments.} \textcolor{black}{Beyond simulated character animation, extending HIL to real-world humanoid systems is also a promising direction. Such applications would require addressing challenges including sim-to-real transfer, perceptive whole-body control, and robustness to real-world dynamics and sensing noise.}


\bibliographystyle{ACM-Reference-Format}
\bibliography{sample-bibliography}


\clearpage

\appendix

In the appendix, we first provide a summary of revisions, highlighting the expanded experiments, and additional baselines. We then present additional details about the controller architecture, training procedure, and evaluation.

\section{Summary of Revisions}
This revised version expands the scope of the paper from parkour-specific skill learning to a more general framework for dynamic athletic control, demonstrating the method’s applicability across both parkour and heading control tasks. We introduce a goal-conditioned observation space that unifies motion tracking and adversarial imitation learning modes. We include additional experiments and more comprehensive baselines: Task Reward, AMP~\cite{DBLP:journals/tog/PengMALK21}, ASE~\cite{DBLP:journals/tog/PengGHLF22}, MaskedMimic~\cite{DBLP:journals/tog/TesslerGNCP24}, and the warm-start variants of Task Reward and AMP. The Heading Task is newly added, providing further evidence of generality of the method. We have also reorganized the Tasks section for clarity, refined the dataset construction and implementation details, and addressed additional feedback from the previous submission.

\section{Controller Representation}

\subsection{Point Cloud Representation vs. Voxel-Based Representation}
Since our obstacles are primarily box-shaped, we sample ~15 points uniformly on the surface of each object, using N=60 points as input for a 180-dimensional representation of nearby obstacles.

We experiment with voxel-based representations following ManipNet. We used a 10x10x10 voxel grid with 20cm spacing per cell. However, this approach performs poorly—even in the motion tracking task, the character struggles to imitate all the reference motions. We find that voxel-based representations introduce a trade-off between perception volume and spatial resolution. In manipulation tasks, objects are small but detailed, allowing fine-grained voxel representations (e.g., 1cm cells) to capture rich object features. However, in human-scene interaction tasks, particularly parkour, objects primarily serve as interaction surfaces rather than complex geometries. A 10x10x10 grid (1000 dimensions) already exceeds our point cloud representation, yet with 20cm cells, the perception volume is limited to 2m x 2m x 2m, losing fine details. Reducing the voxel size would dramatically increase memory requirements, making it impractical to capture larger environments with distant obstacles. Additionally, pointcloud-based representations are widely adopted in other control models, such as 3D Diffusion Policy~\cite{DBLP:conf/rss/ZeZZHWX24} and DexPoint~\cite{DBLP:conf/corl/QinHY0022}.

\subsection{Discriminator Observations}
Our discriminator takes both the character state and the closest $N$ points from the scene as input to evaluate the naturalness of motions within the context of the surrounding environment. To provide sufficient temporal information, we define a state transition as a sequence of the past 10 steps. This design captures richer motion dynamics, enabling the discriminator to assess the overall consistency and flow of movements. For each time step, the following features are included:
\begin{itemize}
    \item Linear velocity and angular velocity of the root, represented
 in the character’s local coordinate frame
    \item Local rotation of each joint
    \item Local velocity of each joint
    \item 3D positions of the end-effectors (e.g. hands and feet), represented in the character’s local coordinate frame
    \item The closest $N$ points from the scene to the root    
\end{itemize}
The root is designated to be the character’s pelvis. The character’s local coordinate frame is defined with the origin located at the root, the x-axis oriented along the root link’s facing direction, and the z-axis aligned with the global up vector. We concatenate the features from 10 consecutive steps to form the input to the discriminator. This design captures the temporal dynamics of the motion, allowing the discriminator to evaluate transitions and interactions. By including character state features and the scene point cloud across these steps, the discriminator gains a richer context to assess the naturalness and scene alignment of the generated motions. For the heading task, we remove the point cloud observation in the discriminator.

\subsection{Actions}
Proportional derivative (PD) controllers are used to actuate each degree of freedom (DoF) in the character's body. For each joint indexed by $i$, the action $a_{t,i}$ specifies the desired joint position, from which the torque $\tau_i$ is computed as: $\tau_i = k_p \cdot (a_{t,i} - q_{t,i}) - k_d \cdot \dot{q}_{t,i}$,
where $q_{t,i}$ and $\dot{q}_{t,i}$ denote the position and velocity of joint $i$ at time $t$. The policy's action distribution, $\pi(a | s, o) = \mathcal{N}(\mu_{\pi}(a | s, o), \Sigma_{\pi})$, is modeled as a multi-dimensional Gaussian. The mean $\mu_{\pi}$ is predicted by the model, and the covariance matrix $\Sigma_{\pi}$ is fixed. Each element on the diagonal of $\Sigma_{\pi}$ is $\sigma_{\pi} = 0.055$, representing the standard deviation of the action outputs for each joint.

\section{Training and Evaluation details}

\subsection{Training Hyper-Parameters}
In the training, the discount factor is set to $\gamma=0.99$. GAE~\cite{DBLP:journals/corr/SchulmanMLJA15} with $\tau = 0.95$ is used. The policy learning rate is $2e-5$ and the learning rate for critic and discriminator is $1e-4$. Training is parallelized with 4096 environments, each having an episode length of 400, distributed across four NVIDIA V100 GPUs. A batch size of 4096 is used for updating the policy, critic, and discriminator on each GPU.

For parkour task, the target reward $r^{target}_t$, we clamp $||p^{root}_{t-1} - g_{t-1}|| - ||p^{root}_{t} - g_{t}||$ to the range $[0, 0.05]$ to prevent the controller from exhibiting unnatural behaviors in an attempt to complete the task as quickly as possible. 

For heading task, we set 
\begin{align}
w^{vel} = 0.7, w^{face} = 0.3, \alpha = 0.25, v^{*} = 1.2.
\end{align}

For the tracking rewards, we set 
\begin{align}
&w_p = 2.5, w_r = 1.5, w_{v} = 0.5, w_{\omega} = 0.5, w_h = 1, , w_{e} = 0.001 \nonumber, \\
&\alpha_p=1.5, \alpha_r = 0.3, \alpha_{v}=0.12, \alpha_{\omega}=0.05,  \alpha_h=20.
\end{align}

We assign a weight of 0.5 to both the tracking/task reward and the style reward.

The transformer utilizes a latent dimension of 256 whereas the internal feed-forward size is 512. It has two layers and two self-attention heads. The encoders for character state and target goal are two MLPs with hidden size [512,256]. The PointNet is a shared MLP with hidden size [512,256]. The critic and discriminator are two MLPs with hidden size [1024,512]. ReLU~\cite{DBLP:conf/icml/NairH10} activations are used for all hidden units.

\subsection{Obstacles}

During training, obstacles are sampled directly from the reference data, forming sequences of five obstacles per course. Adjacent obstacles are spaced 2-3 meters apart, with relative orientations sampled randomly between -20 and 20 degrees. No perturbations are applied during training. However, during evaluation, various noise levels are introduced to assess the controller's robustness.

\subsection{Dynamic Time Warping}
Dynamic Time Warping (DTW)~\cite{DBLP:books/daglib/0019158} is an algorithm designed to measure the similarity or distance between two temporal sequences, even when these sequences are misaligned or have varying speeds. In the context of motion analysis, DTW is used to compute the distance between the reference motion $A$ and generated motion $B$, represented by their respective feature sequences $a = [a_1, a_2, \dots, a_T]$ and $b = [b_1, b_2, \dots, b_{T'}]$, where $T$ and $T'$ are the lengths of the two motions. The core idea of DTW is to align the two feature sequences dynamically, creating an optimal mapping between frames of $A$ and $B$ that minimizes the cumulative distance while respecting their temporal order. For features a and b, we use the same feature for describing the character state in the discriminator.

A cost matrix $C$ is first initialized, where each element $C(i, j)$ represents the pairwise Euclidean distance between features $a_i$ and $b_j$. A warping path $W = [(i_1, j_1), (i_2, j_2), \dots, (i_K, j_K)]$ is then computed through this matrix, defining the optimal alignment between the two motions. The path must satisfy boundary conditions (starting at $(1, 1)$ and ending at $(T, T')$), continuity (each step connects adjacent elements), and monotonicity (time progresses sequentially in both sequences). The cumulative cost along this path, $\sum_{(i, j) \in W} C(i, j)$, represents the DTW distance between $A$ and $B$. In this work, we normalize the DTW distance by dividing it by 1000 to report the tracking error.

Additionally, since the generated motion involves clearing a sequence of obstacles, it is necessary to segment the generated motion into sub-motion clips corresponding to specific obstacle interactions for comparison with the reference motion. To achieve this, we segment the generated motion by comparing the root trajectory of the generated motion with that of the reference motion. Specifically, each sub-motion clip is defined as the segment of the generated motion where the root position is closest to the first and last frames of the corresponding reference motion segment. The generated motion clip is aligned with the relevant obstacle interaction in the reference motion, allowing for a more accurate evaluation of tracking error.

\end{document}